\RequirePackage{fix-cm}
\documentclass[twocolumn]{svjour3}          % twocolumn
\pdfoutput=1
\smartqed  % flush right qed marks, e.g. at end of proof
\usepackage{amsfonts,amsmath,amssymb,marvosym}%mathabx
\usepackage{textcomp}
\usepackage{bm}
\usepackage{graphicx}
\usepackage[latin1]{inputenc}
\usepackage{rotating}
\usepackage{color}
\definecolor{hipsterblue}{rgb}{0, 0.447, 0.741}
\definecolor{hipsterlightblue}{rgb}{0.301, 0.745, 0.933}
\definecolor{hipsterred}{rgb}{0.85, 0.325, 0.098}
\definecolor{hipsteryellow}{rgb}{0.929, 0.694, 0.125}
\definecolor{hipsterbordeaux}{rgb}{0.635, 0.078, 0.184}
\usepackage{upgreek}
\usepackage[urlcolor=blue,colorlinks=true,pdfborderstyle={/S/U/W 1}]{hyperref}
\usepackage[square,numbers,comma,sort&compress]{natbib}
\usepackage{enumerate} 
\usepackage[rightcaption]{sidecap}  %outercaption left/right depending if odd/even pages
\sidecaptionvpos{figure}{t}

\newcommand{\ie}{{i.e.}}
\newcommand{\eg}{{e.g.}}
\newcommand{\etal}{\textit{et~al.}}

\newcommand{\Ap}{A_\mathrm{p}}
\newcommand{\Api}{A_{\mathrm{p}}^{(i)}}
\newcommand{\Ai}{A_I}
\newcommand{\Cm}{C_\mathrm{m}}
\newcommand{\imconc}{N_\mathrm{S}}
\newcommand{\imconcmeas}{N_\mathrm{S}'}
\newcommand{\dmed}{d_\mathrm{med}}
\newcommand{\diamp}{d_\mathrm{p}}

\newcommand{\errz}{\errzglobal^\delta}

\newcommand{\errzj}{\epsilon_{z_i}}
\newcommand{\errxglobal}{\sigma_x}
\newcommand{\erryglobal}{\sigma_y}
\newcommand{\errzglobal}{\sigma_z}

\newcommand{\errxy}{\errxyglobal^\delta}
\newcommand{\errxyglobal}{\sigma_{x,y}}
\newcommand{\GDPTprogram}{\emph{DefocusTracker}}
\newcommand{\gradient}{\alpha}
\newcommand{\Ic}{I_\mathrm{c}}
\newcommand{\It}{I_\mathrm{t}^{(i)}}
\newcommand{\Istack}{I_{\mathrm{c}}^{(k)}}%
\newcommand{\Iback}{I_0}%{I_\mathrm{back}}

\newcommand{\Ipattern}{I_\mathrm{pattern}}
\newcommand{\IO}{I_0}
\newcommand{\meanPI}{\mu_\mathrm{p}}
\newcommand{\Ntot}{N_\mathrm{p}}

\newcommand{\Nm}{N'_\mathrm{p}}
\newcommand{\Ncal}{N_\mathrm{cal}}
\newcommand{\noise}{\sigma_I}
\newcommand{\detectedpglobal}{\phi}
\newcommand{\detectedp}{\detectedpglobal^\delta}
\newcommand{\ximage}{X}
\newcommand{\yimage}{Y}
\newcommand{\xmeas}{x'}
\newcommand{\ymeas}{y'}
\newcommand{\zmeas}{z'}
\newcommand{\xmeasj}{x'_i}
\newcommand{\ymeasj}{y'_i}
\newcommand{\zmeasj}{z'_i}
\newcommand{\xreal}{x}
\newcommand{\yreal}{y}
\newcommand{\zreal}{z}
\newcommand{\Vmeas}{V}

\newcommand{\beq}[1]{\begin{equation} \eqlab{#1}}
\newcommand{\eeq}{\end{equation}}
\newcommand{\bsub}{\begin{subequations}}
\newcommand{\esub}{\end{subequations}}
\def\bal#1\eal{\begin{align}#1\end{align}}
\def\bsubal#1\esubal{\bsub \begin{align}#1\end{align} \esub}

\newcommand{\eqlab}[1]{\label{eq:#1}}
\newcommand{\equref}[1]{Eq.~(\ref{eq:#1})}

\newcommand{\figref}[1]{Fig.~\ref{fig:#1}}

\newcommand{\figlab}[1]{\label{fig:#1}}

\newcommand{\secref}[1]{Section~\ref{sec:#1}}

\newcommand{\seclab}[1]{\label{sec:#1}}
\newcommand{\tabref}[1]{Table~\ref{tab:#1}}

\newcommand{\tablab}[1]{\label{tab:#1}}  
\begin{document}\sloppy

\title{General Defocusing Particle Tracking: \\Fundamentals and uncertainty assessment}

%\thanks{Grants or other notes about the article that should go on the front page should be placed here. General acknowledgments should be placed at the end of the article.}


\author{Rune Barnkob \and Massimiliano Rossi}


\institute{
Rune Barnkob \at
Heinz-Nixdorf-Chair of Biomedical Electronics\\
Department of Electrical and Computer Engineering\\
Center for Translational Cancer Research (TranslaTUM)\\
Technical University of Munich\\
81675 Munich, Germany\\
\email{rune.barnkob@tum.de}
\and
Massimiliano Rossi \at
Department of Physics\\
Technical University of Denmark\\
DTU Physics Building 309\\
DK-2800 Kongens Lyngby, Denmark\\
\email{rossi@fysik.dtu.dk}
}

% \date{\color{red}\today}
\date{Received: date / Accepted: date}

\maketitle

\begin{abstract}

General Defocusing Particle Tracking (GDPT) is a single-camera, three-dimensional particle tracking method that determines the particle depth positions from the defocusing patterns of the corresponding particle images. %In particular, GDPT uses a supervised learning approach based on a reference set of experimental particle images. 
GDPT relies on a reference set of experimental particle images which is used to predict the depth position of measured particle images of similar shape. 
While several implementations of the method are possible, its accuracy is ultimately limited by some intrinsic properties of the acquired data, such as the signal-to-noise ratio, the particle concentration, as well as the characteristics of the defocusing patterns. GDPT has been applied in different fields by different research groups, however, a deeper description and analysis of the method fundamentals has hitherto not been available. In this work, we first identity the fundamental elements that characterize a GDPT measurement. Afterwards, we present a standardized framework based on synthetic images to assess the performance of GDPT implementations in terms of measurement uncertainty and relative number of measured particles. Finally, we provide guidelines to assess the uncertainty of experimental GDPT measurements, where true values are not accessible and additional image aberrations can lead to bias errors. The data were processed using \GDPTprogram, an open-source GDPT software. The datasets were created using the synthetic image generator MicroSIG and have been shared in a freely-accessible repository.


% 

% \PACS{06.30.Gv \and 47.85.-g \and 07.60.Pb \and 87.64.M-}
% 47.80.Cb Fluid mechanics, Velocity measurements 
% 47.80.−v Fluid mechanics, Instrumentation and measurement methods in fluid dynamics 
% Velocity, measurement of, 06.30.Gv
% Fluid mechanics, applied, 47.85.-g
% Microparticles, optical properties of, 78.66.Vs
% Microscopy, optical, in biophysics, 87.64.M-
% Microscopy, optical, conventional, 07.60.Pb
% Visual imaging, 87.63.L-

%\subclass{MSC code1 \and MSC code2 \and more}
\end{abstract}

\section{Introduction}
\label{intro}

Measurement methods based on the imaging of tracer particles in a flow are standard tools in experimental fluid mechanics. Probably the most representative method is the the Particle Image Velocimetry (PIV), introduced in the mid-1980s, which in its basic configuration allows to measure a two-dimensional, two-component (2D2C) flow field by looking at the displacement of tracer particles illuminated by a thin laser sheet  \cite{adrian1984scattering,willert1991digital}. In the following years, also thanks to the improvement of digital cameras, computers, and image analysis software, a great variety of new techniques derived by PIV has come out, allowing time-resolved, 3D3C measurements, at large or microscopic scale. A good overview can be found in the reference textbook by Raffel~\etal~\cite{raffel2018particle}. When the displacements of individual particles are measured, rather than the average particle displacement in interrogation windows, the method is more properly referred to as Particle Tracking Velocimetry (PTV).

Methods derived from PIV or PTV are lately becoming more and more important, not only in experimental fluid mechanics, but also in other disciplines such as medicine, biology, or bio-engineering, in which the experimental characterization of complex fluidic systems, like blood vessels or biochemical microfluidic platforms, is crucial. In this domain, a major role is played by single-camera, 3D PTV methods, which are needed in environments where the flow is three-dimensional and only one optical access, typically through a microscope objective, is available~\cite{taute2015high,rossi2017kinematics,barnkob2018acoustically,rossi2019interfacial,liu2019investigation}. In the past years, the development of single-camera, 3D particle tracking methods has become a research field on its own~\cite{cierpka2012particle}. The major challenge here is to obtain the depth information from two-dimensional images of particles. Several principles have been proposed to solve this problem, such as holography~\cite{memmolo2015recent}, light-field cameras~\cite{fahringer2015volumetric,shi2019volumetric}, or image defocusing~\cite{willert1992three,pereira2000defocusing,wu2005three,zhang2008three,van2012non,cierpka2010simple,rossi2014optimization,barnkob2015general}. Defocusing is a particularly attractive approach, since it does not necessarily require the implementation of special optics or cameras. The main idea is to use optical systems with small depth of field, where the degree of defocusing of the particle images is related to the particles' depth positions.

A large variety of 3D particle tracking methods relying on defocusing has been proposed. A first notable implementation was the Defocusing Digital PIV, where a three-pinhole mask was used to more efficiently read-out the defocusing information \cite{willert1992three,pereira2000defocusing}. Other research groups looked at the changes of the radial intensity profiles of axisymmetric particle images~\cite{zhang2008three,van2012non,afik2015robust}. Another method is the Astigmatic PTV, where an astigmatic aberration, introduced by a cylindrical lens, is used to obtain particle image shapes with a characteristic elliptical shape directly related to their depth position \cite{cierpka2010simple,rossi2014optimization}.

All these methods share the same principle: The particle image changes shape in a systematic fashion depending on the particle's depth position. This principle can be generalized to any optics or image type, by constructing a lookup table that maps the particle image shapes with the corresponding depth positions. This approach was introduced by Barnkob~\etal~\cite{barnkob2015general} and is referred to as the General Defocusing Particle Tracking (GDPT). The same concept was developed independently by Taute~\etal~\cite{taute2015high} to track the 3D motion of bacteria using a standard phase-contrast microscope. Both implementations by Barnkob~\etal\ and Taute~\etal\ used the normalized cross-correlation for comparing the target images with the images in the lookup table; however, different image comparison approaches can be used, and neural networks and artificial intelligence are expected to play a significant role in the future~\cite{newby2018convolutional,munoz2019machine}. 

GDPT has been applied in different fields and by different research groups~\cite{qiu2019experimental,volk2018size,rossi2017kinematics,barnkob2018acoustically,boyko2020nonuniform}. However, the description of GDPT is limited to the seminal papers in Refs.~\citenum{barnkob2015general} and \citenum{taute2015high}, and there is a need for a more thorough definition and analysis of the fundamental elements characterizing a GDPT measurement. This step is crucial for a deeper understanding of the method and for future developments and improvements. Moreover, to be able to compare different GDPT algorithms and implementations, standardized assessment schemes and a database of reference images must be established, in analogy to what is already available for PTV or PIV~\cite{kahler2016main}. Consequently, the objective of this work is to fill this gap and provide the basis for further understanding and development of the GDPT method. 

In \secref{GDPTbasics} we identify the fundamental elements that characterize a GDPT measurement and provide a standardized scheme to assess its uncertainty. Following, in \secref{results_synthetic} we study the uncertainty of the method as a function of image noise and particle image density using synthetic images. Finally, in \secref{biascorrection} we provide guidelines to assess the uncertainty of GDPT measurements in experimental cases where bias errors are present and the true values are not accessible. The synthetic datasets used in this work are presented in Appendix~\ref{sec:datasets} and are freely-available to the research community. The datasets are analyzed using \GDPTprogram, an open-source GDPT implementation described in Appendix~\ref{sec:GDPTalgorithm}.

\begin{figure*}[t!]
\centering
\includegraphics[width=1.0\textwidth]{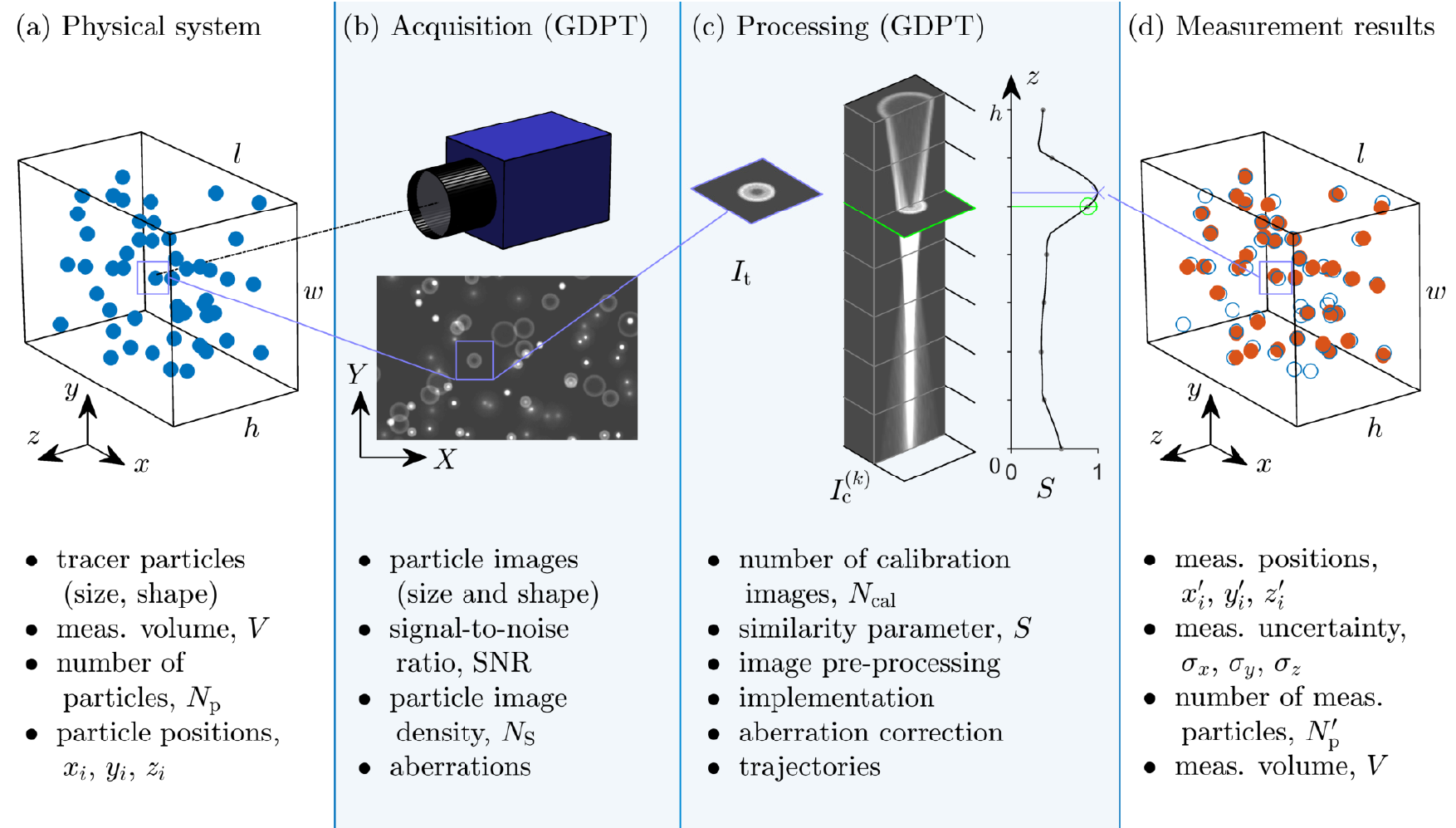}	
\caption{Fundamentals of a GDPT analysis. (a) The analyzed system consists of a measurement volume $\Vmeas = l\times w\times h$ filled with $\Ntot$  particles. (b) The particles in the volume are recorded on one image using a single-camera acquisition approach. As a consequence of image defocusing, the particle images have different shapes according to their depth positions. (c) The depth position of a target particle image is identified using a reference set of calibration images that maps the particle image shapes as a function of their depth position. A similarity function is used to match the target particle image with the most similar calibration image. (d) The result of a GDPT measurement is a set of measured particle coordinates ($\xmeas$,$\ymeas$,$\zmeas$). The accuracy of the measurement result is quantified by the respective estimated uncertainties ($\errxglobal$,$\erryglobal$,$\errzglobal$).}
\figlab{figure-fundamental}
\end{figure*}

\section{Fundamentals of GDPT}
\seclab{GDPTbasics}

The fundamental principles of a GDPT analysis are outlined in \figref{figure-fundamental} and involve (a) the physical system under investigation, (b) a single-camera acquisition approach, (c) an image processing approach, and (d) the evaluation of the measurement results. The list of symbols and parameters used in this section is given in \tabref{symbols}.

\subsection{Physical system}

The purpose of a 3D particle tracking system is to locate the physical coordinates %
\bal
    \xreal_i, \yreal_i, \zreal_i, \quad \text{for} \,\,\, i=1,\ldots,\Ntot,
\eal
of a number $\Ntot$ of tracer particles inside a measurement volume $\Vmeas$ and to track their displacement in time. The tracking problem will not be discussed here; in GDPT analysis, the particle concentration is relatively low and conventional tracking algorithms can be used. The fundamental analysis in this work concerns only the problem of particle location determination in GDPT. As we consider single-camera acquisition, it is convenient to define the reference frame with two in-plane coordinates, $x$ and $y$, perpendicular to the optical axis of the camera objective, and one depth coordinate, $z$, parallel to the optical axis (\figref{figure-fundamental}). The measurement volume can be approximated by a rectangular cuboid with dimensions $l \times w \times h$, being $h$ the dimension in the depth direction. The maximum size in the in-plane direction ($l \times w$) is set by the field of view of the imaging system. The maximum depth $h$ that can be achieved depends on the imaging system, the particle size, and the illumination intensity, and corresponds to the region where the signal of defocused particle images is strong enough to be processed by the image analysis method.

\begin{SCfigure*}
\caption{(a) The particle image function $\Ic$ describe the image of one particle as a function of its depth position. (b) The SNR and particle image area $\Ap$ depends on the particle depth position. (c-d) Given the same number of particle per pixels (ppp), the number of overlapping particles is strongly affected by the particle image size.}
\figlab{figure-general-concepts}
\includegraphics[width=1.5\columnwidth]{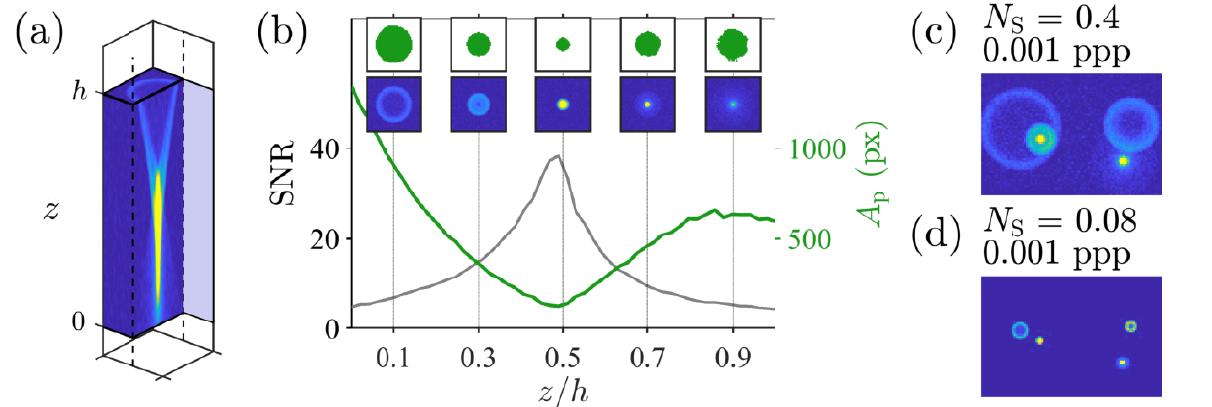}
\end{SCfigure*}

\subsection{Acquisition (GDPT)}
\seclab{acquisition}

When using single-camera systems, the tracer particles in the measurement volume are recorded on a single image. For a given optical system, we can define a particle image function $\Ic(\ximage,\yimage,\xreal,\yreal,\zreal)$ which provides the image of one particle depending on its position in the physical space. $\Ic$ can be seen as a point spread function defined for one particle, where $\ximage$ and $\yimage$ are the coordinates in the image space, and where we impose that the center of the particle image is located at $(\ximage,\yimage) = (0,0)$ for any $\xreal$, $\yreal$, and $\zreal$. 

We consider an idealized optical system with no distortions, and where $\Ic$ is independent of the in-plane coordinates (\figref{figure-general-concepts}(a))
\bal
\label{icapprox}
\Ic(\ximage,\yimage,\xreal,\yreal,\zreal)\approx\Ic(\ximage,\yimage,\zreal). 
\eal
Under this approximation, we can represent a general image $I(X,Y)$, containing a total number $\Ntot$ of particles, as
\begin{equation}
\label{im}
I(\ximage,\yimage) = \sum_i \Ic(\ximage-M\xreal_i,\yimage-M\yreal_i,z_i)+\Iback(\ximage,\yimage),    
\end{equation}
where $M$ is the magnification of the optical system and where $\Iback$ is a term that accounts for background intensity and thermal noise of the camera.

With respect to the final result of a GDPT evaluation, two primary parameters must be considered:
\begin{itemize}
  \item \textbf{Signal-to-noise ratio (SNR)}. Following a convention used in image analysis, we define  the SNR as the ratio between the mean particle image signal $\meanPI$ and the standard deviation of the noise $\noise$
  \bal
  \mathrm{SNR} = \frac{\meanPI}{\noise}.
  \eal
  Practically, $\noise$ can be estimated as the standard deviation of the image intensity in a region without particles, and $\meanPI$ can be calculated as the average intensity of the particle image minus the average background intensity. We can define the boundary of a particle image by setting a threshold, as shown in \figref{figure-general-concepts}(b).
 
  \item \textbf{Particle image density $\boldsymbol{\imconc}$.} This parameter is often given in particles per pixels (ppp). This approach, however, does not consider the particle image size, which is an important factor in defocusing applications with respect to overlapping particles, as shown in \figref{figure-general-concepts}(c-d). For GDPT analysis, it is more convenient to consider the ratio between the sum of the particle image areas $\Api$ and full image area $\Ai$ \cite{kahler2012uncertainty}
  \bal
  \eqlab{imconc}
  \imconc = \frac{1}{\Ai}\sum_{i} \Api \approx  \Ntot\frac{\bar{A}_\mathrm{p}}{\Ai}.
  \eal
  Here, $\bar{A}_\mathrm{p}$ is the average of the particle image areas, which can be estimated for instance from the calibration stack. One can easily translate the $\imconc$ value in ppp, by dividing it with $\bar{A}_\mathrm{p}$. It should be noted that this parameter is equivalent to the source density encountered in classical PIV literature \cite{adrian1984scattering,westerweel2000theoretical}.
\end{itemize}

In real optical systems, the approximation in Eq.~\eqref{icapprox} may not hold as a consequence of optical aberrations. The most common aberrations which are relevant for GDPT applications are:

\begin{itemize}
  \item \textbf{Parallax or perspective error.} The magnification is not constant across the measurement depth, i.e. objects closer to the lens appear larger on the image. This error is typically small and can be neglected for microscope objective lenses.
 \item \textbf{Field curvature.} The focal plane is not flat, therefore the measured $z$ position must be corrected depending on the particle in-plane position. This effect is relevant in GDPT applications since a field curvature of a few micrometers can have a significant impact in the measurement.
 \item \textbf{Distortion.} The particle images are distorted as they move away from the image center. This error  is difficult to correct in GDPT analysis based on a single calibration stack but is normally not strong in conventional optical setups.  
\end{itemize}
More details about practical strategies to deal with optical aberrations and bias errors are discussed in \secref{biascorrection}.

\subsection{Processing (GDPT)}
\seclab{GDPTprocess}

The aim of a GDPT processing is to determine the 3D particle positions from the defocused particle images. The in-plane position can be obtained using one of the many approaches developed for 2D PTV \cite{maas1993particle,ohmi2000particle,ouellette2006quantitative,raffel2018particle}. The out-of-plane component is determined using a reference set of calibration images and must contain the following elements:
\begin{enumerate}
    \item A discrete reference set of calibration images, referred to as the calibration stack.  This can be seen as a discrete sampling at known positions of the particle image function $\Ic$
    \bal
        \Istack(X,Y) = &\Ic(X,Y,z_k) + \Iback(X,Y) \nonumber\\
        &\quad\mathrm{with} \,\, k = 1,2,...,\Ncal.
    \eal
$\Istack$ is typically obtained experimentally by taking subsequent images of a reference particle which is displaced at known positions, for instance using a motorized focusing stage~\cite{barnkob2015general}.
    \item A function or procedure to identify target particle images $\It$ inside the image $I$. This normally relies on segmentation algorithms that can be applied on raw or filtered/pre-processed images. 
	\item A function or procedure to quantify the similarity between different images. This is used to rank the calibration images in $\Istack$ with respect to their similarity to a given target particle image $\It$.
    \item A function or procedure to estimate the final depth position $\zmeasj$ of the target particle from the similarity values between $\It$ and $\Istack$. It should be noted that a simple identification of the most similar image in the stack would produce a discrete output. Interpolation schemes are needed to obtain a continuous output with a ``sub-image'' resolution, in analogy with what is typically done in digital PIV evaluations to obtain sub-pixel resolution. 
\end{enumerate}

More generally, the determination of the depth position in GDPT can be seen as a supervised learning problem, where the calibration stack is the training set used to calibrate a prediction algorithm. The scheme of the prediction algorithm remains the same but it can be trained on different setups just using different calibration stacks. We used in this paper the implementation based on the normalized cross-correlation proposed in Ref. \citenum{barnkob2015general}, but other approaches using neural networks or more sophisticated classification schemes would be natural improvements of the method.

\begin{table}[t!]
\centering
\caption{List of symbols and parameters.}
\begin{tabular}{l l}
\hline\hline
Symbol                          &   Description                 \\ \hline\hline
$\Ai$                           &   Image area                         \\
$\Ap$                           &   Particle image area \\
$\Cm$                           &   Normalized cross-correlation maximum \\
$\dmed$                         &   Median filter size    \\
$I(\ximage,\yimage)$            &   Image \\
$\Iback(\ximage,\yimage)$       &   Image background \\
$\Ic(\ximage,\yimage,\zreal)$   &   Particle image function     \\
$\Istack(\ximage,\yimage)$      &   Calibration image stack   \\
$\It(\ximage,\yimage)$          &   Target particle image \\
$M$                             &   Magnification \\
$\meanPI$                       &   Mean particle image signal \\
$\Ncal$                         &   Number of calibration images  \\
$\Ntot$                         &   Number of particles per image \\
$\Nm$                           &   Number of meas. particles per image \\
$\imconc$                       &   Particle image density (source density)  \\
$\imconc^*$                     &   Critical particle image density \\
$\imconcmeas$                   &   Measured particle image density  \\
$\errxyglobal$                  &   Meas. uncertainty, in-plane coords. \\
$\errxy(z)$                     &   Local meas. uncertainty, in-plane coords. \\
$\errzglobal$                   &   Meas. uncertainty, depth coord.\\
$\errz(z)$                      &   Local meas. uncertainty, depth coord. \\
$\noise$                        &   Image noise level (standard deviation) \\
$\detectedpglobal$              &   Relative num. of meas. particles \\
$\detectedp(z)$                 &   Local relative num. of meas. particles\\
SNR                             &   Image signal-to-noise ratio  \\
$S$ / $S(\cdot,\cdot)$          &   Similarity parameter / function  \\
$\Vmeas=l\, w\, h$              &   Measurement volume  \\
$\xreal$, $\yreal$, $\zreal$    &   Coordinates in physical space    \\
$\xreal_i$, $\yreal_i$, $\zreal_i$    &   Particle coordinates   \\
$\xmeas_i$, $\ymeas_i$, $\zmeas_i$    &   Measured particle coordinates \\ 
$\ximage$, $\yimage$            &   Coordinates in image space \\
\hline\hline
\end{tabular}
\tablab{symbols}
\end{table}

\subsection{Measurement results}
\seclab{assessment}

The result of a GDPT evaluation on a single image, containing a total number $\Ntot$ of particles, is a set of measured particle coordinates $\xmeasj$, $\ymeasj$, $\zmeasj$, where
\bal
    \zmeas_i = \begin{cases}
    \zreal_i + \errzj   & \quad\text{if measured}\\
    \text{undefined}    & \quad\text{otherwise}
\end{cases},
\eal
where $\errzj$ is the measurement error and the same definition applies for the other coordinates. To assess the performance of a GDPT measurement, the following parameters must be considered:
\begin{enumerate}
    \item The measurement uncertainty in the particle position determination.
    \item The relative number of measured particles.
    \item The depth of the measurement volume.
\end{enumerate}
For a fair assessment of a GDPT measurement, all these three parameters should be considered, since they are interconnected among each other. For instance, a stricter validation criterion can reduce the error but at the expenses of a smaller number of measured particles.

\begin{SCfigure*}
\caption{Concept of similarity and discrete sampling of the particle image function. (a) Similarity $S$ between the two images at coordinates $z_1$ (blue) and $z_2$ (red) and the full particle image function along $z$. (b) Illustration of self-similarity, the average similarity between  particle images at identical height but different in-plane positions, as a function of image subpixel displacements and noise. (c) Examples of the calibration stack $\Istack$, the discrete representation of the particle image function $\Ic$. (d) Similarity between neighbor images in calibration stacks of $\Ncal = 15$, 50, or 500 calibration images.}
\figlab{figure-similarity}
\includegraphics[width=1.5\columnwidth]{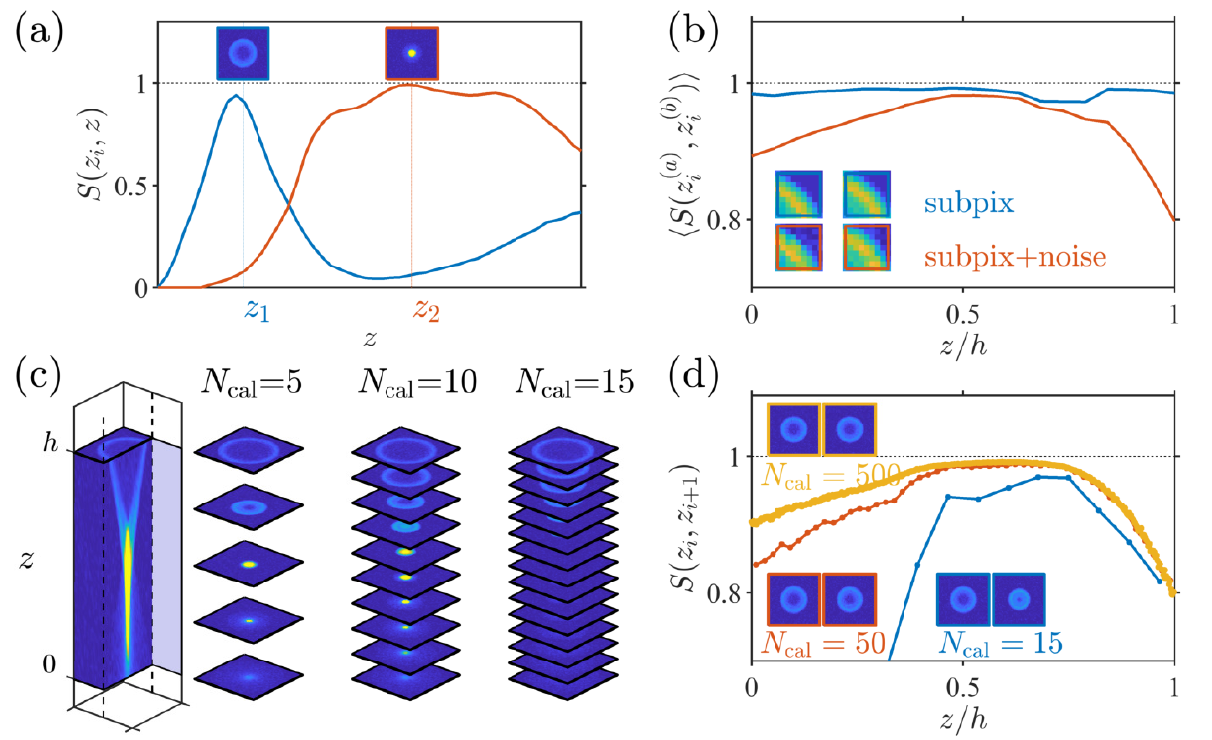}
\end{SCfigure*}

\begin{itemize}
\item \textbf{Measurement uncertainty.}
When the true particle position is known, as in the case of synthetic images, the measurement uncertainty in the determination of one coordinate is normally given in terms of the root-mean-square of the error
\begin{equation}
    \errzglobal =  \sqrt{\frac{\sum_{i}' (\zmeas_i - \zreal_i)^2}{\sum_{i}' 1}} ,
    \label{global_uncertainty}
\end{equation}

where $\sum_{i}'$ indicates summation over the measured values and $\sum_{i}' 1$ is equal to the total number of measured particles $\Ntot'$. Since the particle images have different shapes, a non-uniform uncertainty along $z$ is expected and it is useful to define a local uncertainty 
\begin{equation}
    \errzglobal^{\delta}(z) =  \sqrt{\frac{\sum_{i}' (\zmeas_i - \zreal_i)^2\, \theta^{\delta}(z,\zreal_i)}{\sum_{i}' \theta^{\delta}(z,\zreal_i)}} ,
\label{local_uncertainty}
\end{equation}
with
\bal
    \theta^\delta(z,\zreal_i) = \begin{cases}
    1  & \text{for } \,\, z-\delta < \zreal_i < z+\delta \\
    0               & \text{otherwise}
\end{cases}.
\eal

Eq.~\eqref{local_uncertainty} indicates the uncertainty associated to a bin centered at $z$ and having a width of $2 \delta$. The same applies for the uncertainty of the in-plane coordinates $x$ and $y$. For particle images with rotational symmetry, we can assume  $\errxglobal = \erryglobal$. Note that Eqs.~\eqref{global_uncertainty} and \eqref{local_uncertainty} give the total error, since the true values $\zreal_i$ are used. If the uncertainty is calculated around a mean value, as in case of experimental images, the uncertainty does not take into account bias errors. A discussion about how to apply these formulas to experimental images is presented in \secref{biascorrection}.

\item \textbf{Relative number of measured particles}. This is defined as the ratio between the number of measured particles and the total number of particles in one image
\begin{equation}
\detectedpglobal = \frac{\sum_{i}' 1}{\sum_{i} 1} = \frac{\Ntot'}{\Ntot}.% = \frac{\imconc'}{\imconc}.
\end{equation}
Also in this case it is useful to define a local $\detectedpglobal$ defined as
\begin{equation}
\detectedpglobal^{\delta}(z) = \frac{\sum_{i}' \theta^{\delta}(z,\zreal_i)}{\sum_{i} \theta^{\delta}(z,\zreal_i)}.
\end{equation}
The relative number of measured particles $\detectedpglobal$ tends to decrease as the particle concentration is increased, due to the more frequent occurrence of overlapping particle images. For a given GDPT implementation, there will be a critical particle image density $\imconc^*$ above which the number of measured particles $\Nm$ starts to decrease. This sets the maximum possible seeding density that should be used for that implementation. Furthermore, it should be noted that $\detectedpglobal$ is not directly accessible in experimental images, where only the number of measured particles $\Ntot'$ is available.

\item \textbf{Measurement volume.}
The choice of the depth $h$ of the measurement volume affects the total number of particles $\Ntot$ to be measured and, consequently, the number of overlapping particles and the measurement uncertainty. For instance, a large depth $h$ can include highly-defocused particle images, which have a smaller SNR and are more difficult to detect. To fairly estimate the uncertainty of a given implementation, the measurement depth on which it is applied must be indicated. In GDPT systems, $h$ is practically set by the lowest and highest $z$ coordinate in the stack. 
\end{itemize}

\begin{SCfigure*}
\caption{GDPT uncertainties as a function of calibration sampling and image signal-to-noise ratios [Dataset I, \figref{figure-datasets}(a)]. (a,c,e) Local depth coordinate uncertainty $\errz(z)$ as a function of the depth coordinate $z$ for (a) noise-less images ($\noise=0$) and varying number of calibration images $\Ncal$, (c) fixed number of calibration images $\Ncal=50$ and varying noise level $\noise$, and (e) fixed number of calibration images $\Ncal=50$, fixed noise level $\noise=50$, and for the application of different median filters $\dmed$ to the images prior to the GDPT analysis. (b,d,f) Depth coordinate uncertainty $\errzglobal$ as a function of the number of calibration images $\Ncal$ for (b) noise-less images $\noise=0$ with and without sub-image interpolation, (d) varying noise level $\noise$, and for (f) fixed noise level $\noise=50$ and use of different image median filtering $\dmed$.}
\figlab{figure-grid}
\includegraphics[width=1.5\columnwidth]{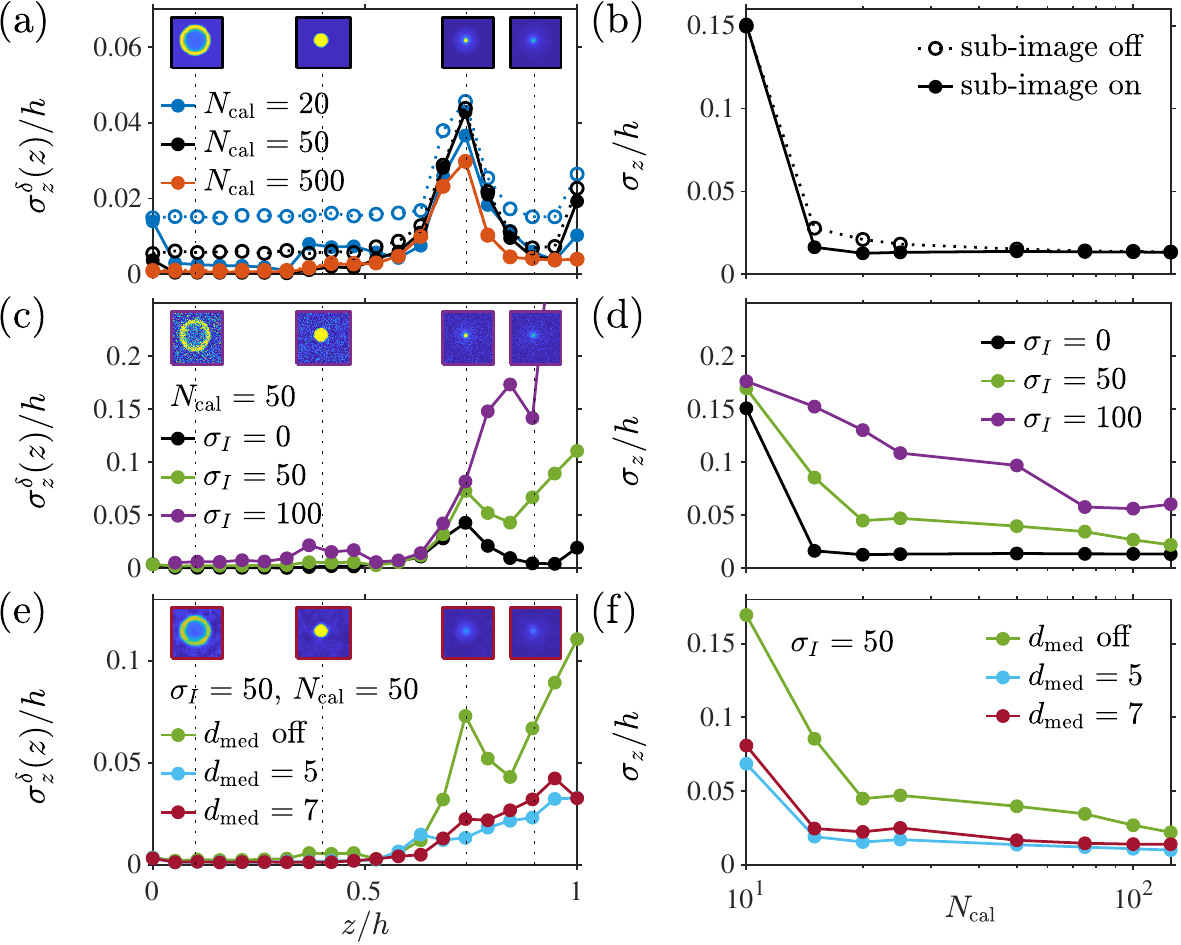}
\end{SCfigure*}

\subsection{Concept of similarity}
\seclab{similarity}

The calibration stack $\Istack$ represents a discrete sampling of the particle image function across $z$. The depth position of a particle is obtained by comparing a target particle image $\It $ with the images in the stack using a similarity function $S$, usually normalized between 0 and 1, with 1 being the perfect match  (\figref{figure-similarity}(a)). In an ideal case, the measurement resolution can be increased indefinitely by increasing the number of images $\Ncal$ in the stack (\figref{figure-similarity}(c)). In a real case, however, beyond a certain $\Ncal$ the difference in shape between two neighbor images will be smaller than the difference induced by the image noise and in-plane image discretization (i.e. the light-intensity distribution is discretized into pixels).

This concept is illustrated in \figref{figure-similarity}(b), where the self-similarity, i.e. the average $S$ between particle images at the same height but different in-plane positions, is shown as a function of $z$. Even in the case with no noise (blue line), $S$ is not unity due to tiny differences induced by sub-pixel displacements of the particle images. By adding noise (red line, $\noise = 25$),  $S$ decreases significantly, especially in regions where the SNR is lower. 
A way to check whether a calibration stack is over-sampling $\Ic$, is to calculate the similarity between neighbor images in the stack, i.e., $S(\Ic^{(k)},\Ic^{(k+1)})$. In the example in \figref{figure-similarity}(d), the stack with $\Ncal = 500$ (orange line) is clearly over-sampled, since it coincides with the self-similarity curve in \figref{figure-similarity}(b). Further refining the stack will not bring any advantage. For $\Ncal = 15$ (blue line), the $S$ between neighbor images is small, showing that the stack under-samples the particle image function. The stack with $\Ncal = 50$ (red line) is a better compromise, with a slightly under-sampled region for $z/h < 0.5$ and over-sampled above.

\section{Standardized uncertainty assessment using synthetic images}
\seclab{results_synthetic}

For a complete assessment of the effect of $\Ncal$, SNR, and $\imconc$ on the final results, it is necessary to look at the final error in the $z$ determination, which is also affected by the choice of the similarity function and the practical implementation of the method (\eg\, algorithms used, interpolation schemes, and smoothing). In this section, we present a systematic procedure based on synthetic images to address these aspects.  Guidelines for the uncertainty assessment in experimental cases is provided in the next section. 

The particle images are generated using MicroSIG~\cite{rossi2019synthetic}. We consider here three datasets to analyze, respectively, the effect of noise, particle concentration, and background intensity gradients. The datasets are freely available\footnote{The datasets can be downloaded through \href{https://defocustracking.com/}{https://defocustracking.com/}.}, more details are given in Appendix~\ref{sec:datasets}. The datasets are analyzed using \GDPTprogram\, which is an open-source GDPT implementation. As similarity parameter, \GDPTprogram\ uses $\Cm$, the maximum of the normalized cross-correlation between two images; for more details, see Appendix~\ref{sec:GDPTalgorithm}.

\begin{SCfigure*}
\caption{GDPT uncertainties as a function of particle image density and overlapping particle images [Dataset II, \figref{figure-datasets}(b)]. (a) The local depth coordinate uncertainty $\errz(z)$, (c) the local in-plane coordinate uncertainty $\errxy(z)$, and (e) the local relative number of measured particles per image $\detectedp(z)$ as a function of the depth coordinate $z$ for particle image densities $\imconc=0.05$ (squares) and $\imconc=0.59$ (circles). (b,d) The uncertainties $\errzglobal$ and $\errxyglobal$, and (f) the relative number of measured particles per image $\detectedpglobal$ (blue colors) and measured particle image density $\imconcmeas$ (orange colors) as a function of the particle image density $\imconc$.}
\figlab{figure-overlapping}
\includegraphics[width=1.5\columnwidth]{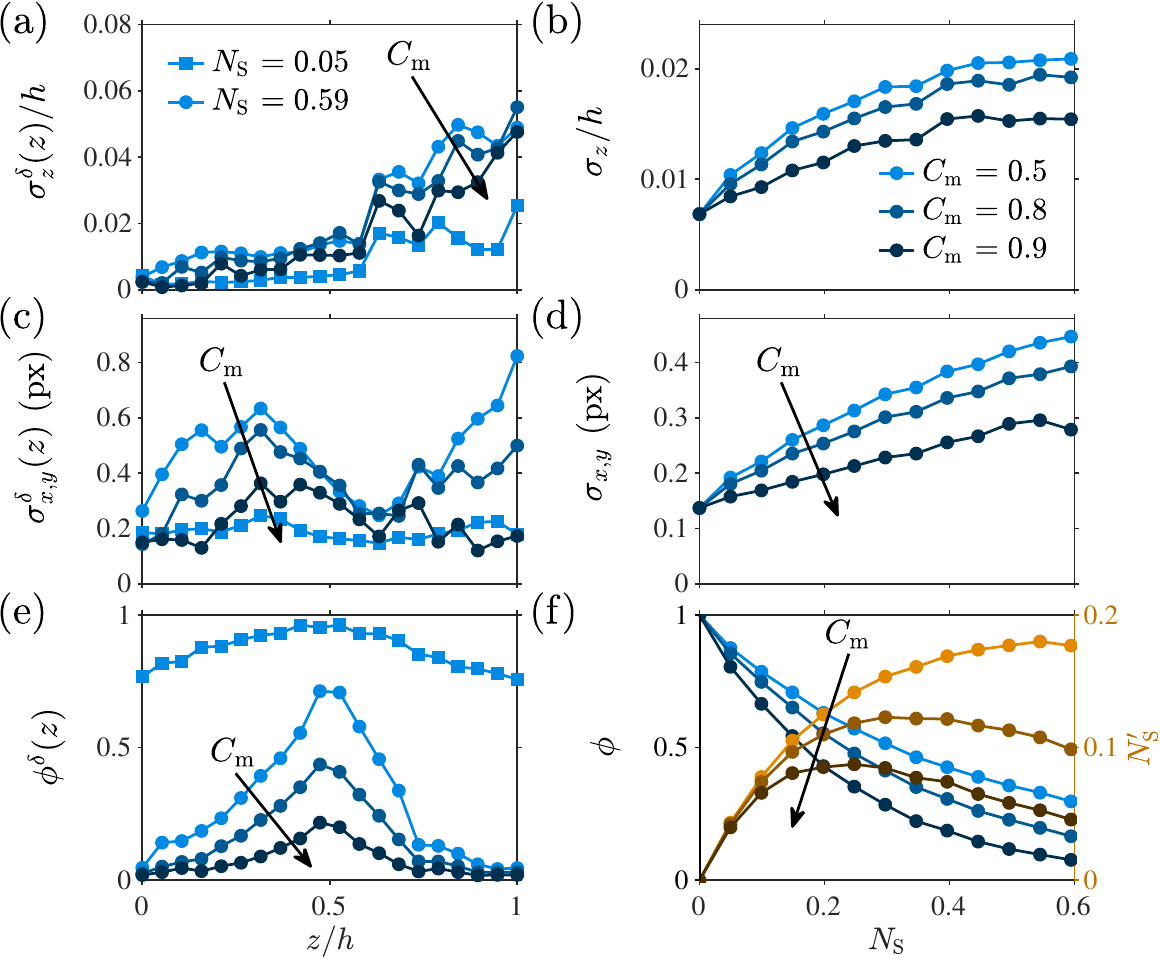}
\end{SCfigure*}

\subsection{Calibration sampling and signal-to-noise ratio}

In this section we evaluate the effect, on the depth coordinate determination, of the number of calibration images and sub-image interpolation along with the influence of image noise. More specifically, we analyze Dataset~I (\figref{figure-datasets}(a) in Appendix~\ref{sec:datasets}) where particle images do not overlap; thus all the particles are detected, i.e., $\Ntot'= \Ntot$.

To isolate the different effects, we first analyze  images with no noise, \ie\ $\noise=0$. The results are shown in \figref{figure-grid}(a), where we plot the local depth coordinate uncertainty $\errz(z)$ as a function of $z$ along with a selection of the corresponding particle image shapes. The results are shown when using $\Ncal = 15$, 50, or 500 calibration images with (points) and without (open circles) the use of sub-image interpolation. It may be noted that the defocused particle images have a sharp ring shape below the focal position (low $z$) and blurred Gaussian-like shape above (high $z$). This is due to the spherical aberration which is present in many optical systems and included in MicroSIG~\cite{rossi2019synthetic}. Detecting the shape differences for blurred particle is more difficult, which explains the larger uncertainty in regions with $z/h>0.6$.

Generally, increasing $\Ncal$ leads to a decreasing local uncertainty $\errz$, but \eg\ for $z/h\sim 0.6$, the use of a sub-image detection scheme can lead to better results for lower $\Ncal$ which is seen by a better performance for $\Ncal=20$ (blue lines) than for $\Ncal= 50$ (black lines) or $\Ncal= 500$ (red lines). The use of a sub-image interpolation scheme allows for a continuous $z$ determination and does in general yield lower uncertainties up to a certain value of $\Ncal$. This is shown in \figref{figure-grid}(b), where we show the depth coordinate uncertainty $\errzglobal$ as a function of $\Ncal$. As $\Ncal$ approaches 200, the uncertainty, with and without the use of a sub-image scheme, converges to the same value.

We now fix the number of images in the calibration stack to $\Ncal=50$ and analyze images with increasing noise levels. The results are presented in \figref{figure-grid}(c). Not surprisingly, the local depth coordinate uncertainty $\errz$ increases with a decreasing SNR. As explained before, the local uncertainty for larger values of $z$ is larger due to the blurred Gaussian-like shape of the particle images. The effect of noise on the local uncertainty naturally propagates into the total uncertainty as seen in \figref{figure-grid}(d). Clearly, as the noise level increases, a larger $\Ncal$ is needed in order to reach a converging uncertainty $\errzglobal$.

Lastly, to explore the reduction of noise through image pre-processing, we now fix the noise level to $\noise=50$ and apply different median filters to the images. We illustrate this in \figref{figure-grid}(e) where we apply a median filter of $\dmed\times\dmed$ (with $\dmed$ equal to 5 or 7) to both the calibration and the  measurement images. The median filter decreases the impact of noise in the local uncertainty $\errz$, in some region even as much as one order of magnitude. Evidently, the filter size has an optimum between averaging out noise and actual particle image features --- this is seen by the presence of some slightly lower values of $\errz$ for $\dmed=5$ than for $\dmed=7$. This is confirmed in \figref{figure-grid}(f), where the same trend is seen for the average uncertainty $\errzglobal$.

\begin{SCfigure*}
\caption{GDPT uncertainties as a function of variations in image background intensities [Dataset~III, \figref{figure-datasets}(c)]. (a) The local depth-coordinate uncertainty $\errz(z)$ and (c) the local relative number of measured particles per image $\detectedp(z)$ as a function of the depth coordinate $z$ for an applied linear background intensity gradient $\gradient=0$ (blue lines), $\gradient=5$ (red colors), and $\gradient=10$ (yellow colors). (b) The depth-coordinate uncertainty $\errzglobal$ and (d) the relative number of measured particles per image $\detectedpglobal$ as a function of $\gradient$.}
\figlab{figure-overlapping-gradient}
\includegraphics[width=1.5\columnwidth]{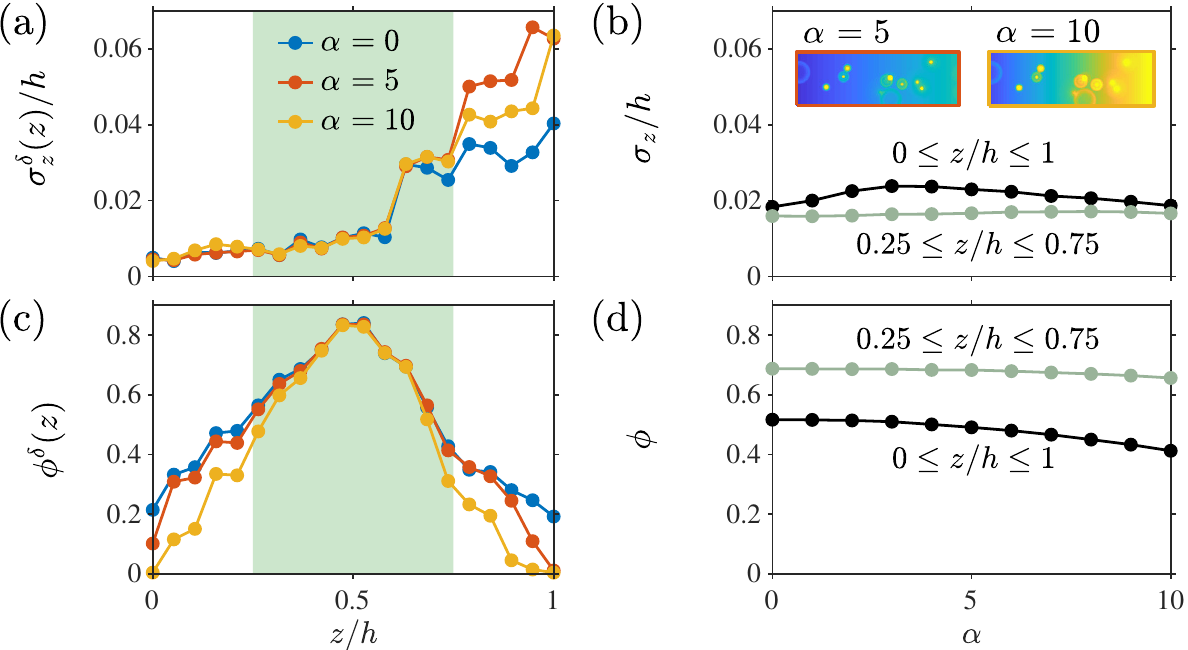}
\end{SCfigure*}

\subsection{Particle image overlap and particle concentration}

After investigating the effect of noise and $\Ncal$, we now analyze the more experimentally-realistic case where the presence of overlapping particle images is possible. In particular we use the noise-less ($\noise=0$) Dataset~II (\figref{figure-datasets}(b)) to assess the GDPT uncertainties as a function of increasing particle concentration while keeping the number of calibration images fixed ($\Ncal=50$). We consider here particle image densities up to $\imconc = 0.59$, corresponding to 0.001 particles per pixels. As the particle concentration increases, the number of overlapping particle images increases concurrently. This leads to lower similarities $\Cm$ between target particle images and calibration images, which furthermore leads to larger errors in the determination of the $z$-coordinates. By setting a higher threshold on the accepted values of $\Cm$, poorly-determined $z$-coordinates can be removed to increase the determination accuracy but with the cost of lowering the number of measured particles $\Nm$.

In Figs.~\ref{fig:figure-overlapping}(a), \ref{fig:figure-overlapping}(c), and \ref{fig:figure-overlapping}(e), we show the local uncertainties $\errz$ and $\errxy$, and the local relative number of measured particles $\detectedp$ for two particle image densities of $\imconc=0.05$ (squares) and $\imconc=0.59$ (circles), respectively.  As $\imconc$ is increased from 0.05 to 0.59, the local errors $\errz$ and $\errxy$ are consistently increasing, while the local relative number of measured particles $\detectedp$ is consistently decreasing.
As the threshold on $\Cm$ is increased, both $\errz$, $\errxy$, and $\detectedp$ are decreasing as expected. The same trend is confirmed in Figs.~\ref{fig:figure-overlapping}(b), \ref{fig:figure-overlapping}(d), and \ref{fig:figure-overlapping}(f), where the uncertainties $\errzglobal$ and $\errxyglobal$, and the relative number of measured particles $\detectedpglobal$ are computed as a function of the particle image density $\imconc$ and $\Cm=0.5$, 0.8, and 0.9, respectively. In addition, we see in \figref{figure-overlapping}(f) that as $\imconc$ increases, the measured particle image density $\imconcmeas$ (orange colors) reaches a peak, \eg,~for $\imconc\approx 0.24$ when $\Cm=0.9$. In fact, as the particle concentration is increased, the number of particles in the image increases but so does the number of outliers or not-detectable particle images due to the increased number of overlapping particle images that cannot be processed. This can be seen as a critical value, denoted here as $\imconc^*$, that sets the maximum particle image density that can be used with those settings.

\subsection{Variations in image intensity}

Experimental measurement images are prone to variations in image light intensity, for example, if the image illumination is inhomogeneous over the image plane or if the illumination amplitude changes over time. This can eventually affect the accuracy of defocusing-based particle detection, but depends on the type of image intensity variation and applied detection algorithm. In this section, we assess the GDPT measurement uncertainties when analyzing measurement images with an applied linear light intensity gradient of increasing steepness. More specifically, we use Dataset~III (\figref{figure-datasets}(c)), which is based on the images in Dataset~II for $\imconc=0.30$ and to which an intensity gradient $\gradient$ has been applied.

Figures~\ref{fig:figure-overlapping-gradient}(a) and \ref{fig:figure-overlapping-gradient}(c) show the local depth coordinate uncertainty $\errz(z)$ and the local relative number of measured particles $\detectedp(z)$. For small $\gradient$, the changes are small as expected from the fact that the normalized cross-correlation is insensitive to changes in intensity level. However, as $\gradient$ increases, the intensity gradient becomes comparable to the intensities of the particle image features. This affects the uncertainty and the number of measured particles specially in regions where the SNR is low (outside the green area in \figref{figure-overlapping-gradient}(a) and (c)). Overall, the effect of intensity gradient is small in the region with large SNR, as shown in \figref{figure-overlapping-gradient}(b) and (d). Note also the previously discussed connection between the uncertainty and number of measured particles, \eg,\ without considering the latter, one would erroneously observe in \figref{figure-overlapping-gradient}(b) a better performance for $\gradient=10$ than for $\gradient=5$.

\begin{SCfigure*}
\caption{Example of uncertainty assessment in experimental GDPT measurement of a Poiseuille flow in a microfluidic channel of rectangular cross-section of 380 $\upmu$m $\times$ 100 $\upmu$m. Through the ansatz of zero cross-sectional velocity, we can estimate the local measurement uncertainties (a) $\errxy$ of the in-plane coordinates and (b) $\errz$ of the depth coordinate.}
\figlab{figure-labchip2015}
\includegraphics[width=1.4\columnwidth]{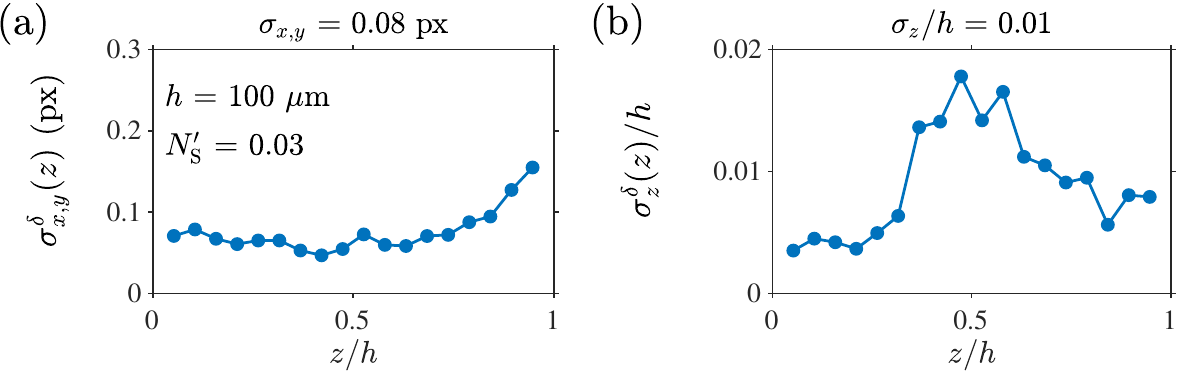}
\end{SCfigure*}

\section{Uncertainty assessment on experimental images}
\seclab{biascorrection}

The fundamental uncertainty assessment presented in this work is only perfectly-applicable to synthetic images. When moving to experimental images, the true values are obviously unknown and additional factors can bias the measurement result. In this section, we briefly discuss the most relevant sources of bias error that should be taken into account and provide guidelines about how to estimate the uncertainty on experimental images.

\subsection{Refractive index}

A common way to obtain the calibration stack experimentally is to use one particle at a fixed position (e.g. stuck to a wall) and to move the objective lens at different depth positions. The reference $z$ coordinates are obtained from the reading of the scanning device. During the measurement, however, the situation is reversed: The objective lens is fixed and the particles move. This has consequences if the immersion medium of the lens is different from the medium where the particles are dispersed, and a correction coefficient must be multiplied to the measured $z$. In most applications, this coefficient is equal to the ratio between the refractive index coefficient of the fluid (typically water) and the one of the immersion medium of the lens (typically air)~\cite{rossi2012effect}.

\subsection{Aberrations}

In real applications, the particle image function $\Ic$ is not the same across in-plane positions due to optical aberrations. The more straightforward way to address this problem would be to take multiple calibration stacks for different in-plane positions; however, this approach is difficult to implement.  First, the size of the calibration stack would increase considerably and so the time necessary for the calibration (a particle must be scanned in different positions). Second, higher complexity and computational costs would be needed to navigate the different calibration stacks and an interpolation scheme must be implemented to cover the entire area of the sensor (only a discrete set of in-plane positions can be taken). Future implementations of the GDPT method using neural network might be able to tackle this task.

In many practically applications, however, the aberrations are weak and a single calibration stack (typically obtained in the center of the image) can be used in the whole image area~\cite{rossi2019interfacial,barnkob2018acoustically,qiu2019experimental}. Additional bias errors due to field curvature or perspective errors can be corrected using a reference experiment, for instance, using a fixed array of tracer particles on a flat plane perpendicular to the optical axis and scan it at different depth positions. 

In microfluidics applications, where microscope objectives are used, the perspective error is typically negligible, but the field curvature can give errors of few microns~\cite{cierpka2010calibration}, which might be relevant at those scales. If the measurements are performed on a straight microchannel, one practical way to correct this bias is by a direct measurement of a Poiseuille flow inside the duct. In this case, we know that the streamlines in such flow must be straight lines, and we can use this information to determine and correct the field curvature.

\subsection{Uncertainty estimation}

To estimate the measurement uncertainty of an experimental GDPT setup, we can use Eqs.~\eqref{global_uncertainty} and \eqref{local_uncertainty}, but we need to estimate the unknown true values. A practical approach can be to measure particle displacements in flows having zero velocity in one direction. We can estimate then the displacement uncertainty, considering 0 as the true value. Modeling the error as normally distributed, the positioning uncertainty will be equal to the displacement error divided by $\sqrt{2}$. Of course, particular care should be taken with reducing as much as possible the bias errors and eliminating false positive with an appropriate outlier rejection scheme. Together with the uncertainty, it is important to indicate the depth $h$ of the measurement volume. When it is possible to estimate $\imconc$, for instance, if the experimental particle concentration is known, the relative number of measured particles $\detectedpglobal$ can be calculated; otherwise, $\imconc'$ should be indicated. 

We give an example of the uncertainty estimation on experimental measurements in \figref{figure-labchip2015} using data from Ref.~\citenum{barnkob2015general}. The data contain GDPT measurements of a Poiseuille flow in a microchannel with a rectangular cross-section of 380 $\upmu$m $\times$ 100 $\upmu$m. The flow is directed along the $x$ direction and the uncertainty was estimated along the perpendicular directions $y$ and $z$, i.e., the directions of zero flow velocity. For a $\imconc' = 0.03$ (very few overlapping particles) and across a height of $h = 100~\upmu$m, we obtained uncertainties $\errxyglobal = 0.08$ px and $\errzglobal/h = 0.01$, in good agreement with the values obtained in the analysis of the synthetic images.

\section{Conclusions}

We have identified the fundamental elements characterizing a GDPT measurement; these are the number of images in the calibration stack $\Ncal$, the image signal-to-noise ratio SNR, the particle image density $\imconc$, and the function chosen to evaluate the similarities between calibration and target images. For the evaluation of GDPT measurements, we presented an assessment scheme encompassing the measured coordinate uncertainties, the relative number of measured particles, and the depth of the measurement volume. 

For a practical implementation of the presented assessment scheme, we created a group of freely-available synthetic image datasets and we used them to study the performance of the GDPT software \GDPTprogram. The fundamental findings of the study are: (\emph{i}) There is an optimal number of images in the calibration stack, including more images will not reduce the measurement uncertainties, (\emph{ii}) the presence of overlapping particle images increases significantly the uncertainty; using a stricter criterion to accept valid measured particles can help to improve the accuracy but at the expense of the number of measured particles, and (\emph{iii}) for each setting, one can extrapolate a critical particle image density beyond which the number of measured particle decreases. The most relevant quantitative results of the study are summarized in \tabref{performance_defocustracker}.

\begin{table}[t!]
\centering
\caption{Summary of the performance on Dataset I (first row) and II (second and third rows) using \GDPTprogram, with  median filter set to $\dmed = 5$. }
\begin{tabular}{l l l l l l l}
\hline
$\Cm$ & $\Ncal$ & $\noise$ & $\imconc^*$  & $\errxyglobal$ (px) & $\errzglobal/h$ 
& $\detectedpglobal$ \\
\hline
0.5 & 50 & 50  & - &   0.152  & 0.014 & 1\\ 
0.5 & 50 & - & 0.545 &   0.436 & 0.021 & 0.330\\ 
0.9 & 50 & - & 0.248 &   0.213 & 0.013 & 0.353\\
\tablab{performance_defocustracker}
\end{tabular}
\end{table}

We provided guidelines to assess the uncertainty of GDPT measurements in experimental cases where bias errors can be present and in contrary to synthetic images, the true values are not accessible. Bias errors occur because of discrepancies between calibration and target images induced by optical aberrations. This should be taken into account, \eg\, by implementing multiple calibration stacks for strong aberrations, or by implementing reference experiments for weak aberrations where one single calibration stack can still be used. To estimate experimental uncertainties, one needs to estimate the true values. This is possible for instance by looking at a velocity flow fields with null velocity in one measurement direction.

In conclusion, this work provides the tools for a more aware use of GDPT and is an important step toward the further development of the method in terms of accuracy, efficiency, and processing speed. Next steps include the expansion of the datasets to temporal-coherent data, non-monodisperse particles, and experimental data, for instance including biological cells for use of GDPT in biomedical sciences.

\begin{acknowledgements}
This work was supported by the European Union's Horizon 2020 Research and Innovation Programme under the Marie Sklodowska-Curie Grant No.  713683  (COFUNDfellowsDTU).
\end{acknowledgements}

\appendix

\begin{figure*}[t!]
\centering
\includegraphics[width=1.0\textwidth]{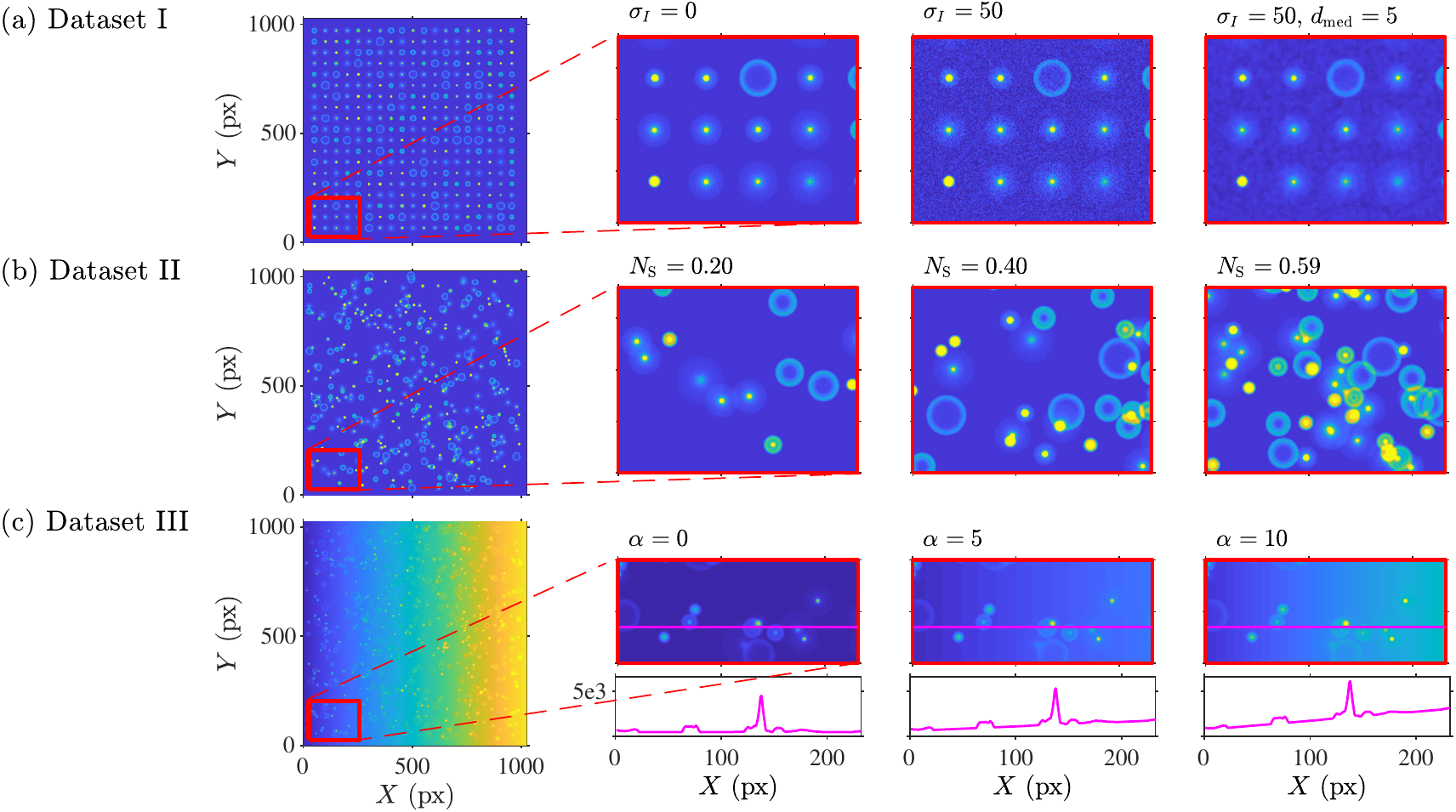}
\caption{Overview of the three synthetically-created datasets used to test and evaluate the performance of GDPT. (a) Dataset I contains measurement images of particles with random $x$, $y$, and $z$ positions but fixed within a grid in $x$ and $y$. The dataset contains groups of measurement images of various noise levels $\noise$, here illustrated via the insets (red rectangles) showing the images for $\noise=0$ and 50, and for $\noise=50$ when a median filter $\dmed=5$ has been applied. (b) Dataset II contains groups of measurement images of randomly distributed particles for different particle image density $\imconc$, here illustrated via the insets (red rectangles) showing the images for $\imconc=0.20$, 0.40, and 0.59. (c) Dataset III contains groups of the measurement images for various added linear light-intensity gradients $\gradient$, see \equref{gradient}. The measurement images are taken from Dataset II for $\imconc=0.30$.}
\figlab{figure-datasets}
\end{figure*}

\section{Datasets for standardized evaluation of GDPT methods}
\seclab{datasets}

In this section, we introduce a group of three datasets and corresponding calibration images for evaluating the performance of a GDPT implementation as a function of different signal-to-noise ratio SNR, particle image density $\imconc$, and background intensity gradients. The datasets do not include the effect of field curvature or image aberrations. The datasets are meant to act as a reference set for the scientific community and are freely available\footnote{The datasets can be downloaded through \href{https://defocustracking.com/}{https://defocustracking.com/}.}. 

The datasets are based on synthetic images created using MicroSIG, which is a Synthetic Image Generator (SIG) using ray tracing and a simplified spherical lens model to obtain realistic defocused or astigmatic particle images~\cite{rossi2019synthetic}. MicroSIG is open source and can be downloaded at \href{https://gitlab.com/defocustracking}{gitlab.com/defocustracking}. In this work, we consider bright, monodisperse particles of diameter $\diamp$, which is the most common case in velocimetry applications. All datasets share the same basic MicroSIG settings in terms of particle diameter ($\diamp$ = 2 $\upmu$m), objective lens (magnification $M$ = 10$\times$,  numerical aperture NA = 0.3, focal length $f$ = 350 $\upmu$m), and sensor settings (pixel size of 6.5 $\upmu$m, $\IO$ = 500 counts). The settings have been chosen since they simulate experimental conditions that are representative for a large number of applications in microfluidics. The simulated measurement depth $h=86~\upmu$m is kept the same for all datasets.

The three datasets are shown in \figref{figure-datasets}. In order to ensure the same level of statistical significance, all datasets contain $\sim$ 20,000 particles for each set of parameters. The three datasets are:
\begin{itemize}
\item \textbf{Dataset I.} This dataset contains $3\times 60$ measurement images for 3 different noise levels $\noise$. Each measurement image contains 361 particle images located at random $x$, $y$, and $z$ positions. The $x$ and $y$ coordinates are loosely constraint on a $19\times 19$ grid to exclude particle image overlapping. The dataset is suitable for analyzing the depth coordinate precision and how it depends on the image noise level $\noise$, number of calibration images $\Ncal$, and method parameters such as image filtering and sub-image approach for the depth coordinate determination.
\item \textbf{Dataset II.} This dataset contains 12 subsets of measurement images of particles with randomly distributed $x$, $y$, and $z$ coordinates, thus including particle image overlapping. Each subset corresponds to a specific particle image density $\imconc$ and contains a certain number of measurement images in order to have an overall number of 20,000 particles. We start with a subset of 1200 images at $\imconc = 0.05$ and end with a subset of 100 images at $\imconc = 0.59$. This dataset is suitable for analyzing the relative number of measured particles and depth coordinate uncertainty as a function of increasing particle image density. 
\item \textbf{Dataset III.} This dataset contains 10 subsets of 34 measurement images of 600 particles ($\imconc=0.30$) with randomly distributed $x$, $y$, and $z$ coordinates. Each subset has a superimposed linear light-intensity gradient along the horizontal direction defined as
\bal
\eqlab{gradient}
 \Ipattern(\ximage,\yimage) = \gradient\ximage,
\eal
where $\gradient$ is a parameter accounting for the gradient intensity. The impact of $\gradient$ can be better appreciated by normalizing its value with the mean particle image intensity divided by the characteristic size of the particle images
\bal
 \tilde{\gradient}=\gradient\,\Ap^{1/2}/\meanPI.
\eal
A value of $\tilde{\gradient}=1$ indicates a light-intensity gradient on the same order of magnitude of the intensity gradient in the particle images. For this dataset we have $\gradient$ ranging from 1 to 10, corresponding to $\tilde{\gradient}$ ranging from 0.16 to 1.6.
\item \textbf{Calibration images.} The set of calibration images provides the calibration images for analyzing the Datasets I--III with GDPT. It contains $3\times 12$ calibration image stacks for 3 different noise levels $\noise$ and 12 different numbers of calibration images $\Ncal$.
\end{itemize}

\section{\GDPTprogram}
\seclab{GDPTalgorithm}

The GDPT implementation used in this work is the Version 1.0 of \GDPTprogram, which is a MATLAB implementation published under the open-source license and available at \href{https://www.defocustracking.com}{defocustracking.com}. The implementation is based on the normalized cross-correlation for image comparison and a polynomial scheme for sub-image interpolation. In order to work fast and robustly, the implementation uses a cross-correlation prediction scheme based on the set of calibration images; for more details, see Ref.~\citenum{rossi2019automatic}. \GDPTprogram\ allows multiple processing iterations using different similarity parameter thresholds to improve the detection of overlapping particles; however, in this work we limit ourselves to one iteration and a single similarity parameter threshold. The implementation allows image noise filter through Gaussian and median filter, though in this work we utilize only the latter. In addition, \GDPTprogram\ allows the determination of particle trajectories through a nearest-neighbor tracking scheme. We do not perform tracking of particles across the image frames in this work, but it is important to mention that using predictive tracking approaches can increase the performance of particle detection as well.

In \GDPTprogram\, the similarity parameter between two images $I_1$ and $I_2$, is defined as the peak maximum of their normalized cross-correlation function $c(u,v)$. Here $u$ and $v$ are the in-plane coordinates in correlation space. Following the seminal paper by Lewis in 1995~\cite{Lewis1995fast}, the normalized cross-correlation takes the form 
\begin{align}
	&c(u,v) = \nonumber\\
	&\frac{\sum_{\ximage,\yimage}[I_1(\ximage,\yimage)-\bar{I_1}]\,[I_2(\ximage-u,\yimage-v)-\bar{I_2}]}{\Big\{\sum_{\ximage,Y}[I_1(\ximage,\yimage)-\bar{I_1}]^2\,[I_2(\ximage-u,\yimage-v)-\bar{I_2}]^2\Big\}^{1/2}},
\end{align}
where $\bar{I_1}$ and $\bar{I_2}$ are the mean image intensities of $I_1$ and $I_2$, respectively. The correlation function $c(u,v)$ has its peak maximum $\Cm$ at the position of best match between the images and the amplitude of the peak maximum ranges from 0 to 1, where 1 indicates a perfect match. A key advantage of the normalized cross-correlation function is that it is robust against light-intensity fluctuations such as inhomogeneous light distribution.


%BibTeX users please use one of
% \bibliographystyle{spphys}        % APS-like style for physics
% \bibliographystyle{ieeetr}
%\bibliography{references}

\end{document}